# Antimony Doping Effect on the Superconducting Properties of SmFeAs(O,F)

Mohammad Azam[1], Manasa Manasa[1], Tatiana Zajarniuk[2], Ryszard Diduszko[3], Tomasz Cetner[1], Andrzej Morawski[1], Andrzej Wiśniewski[2], Shiv J. Singh[1*]

[1]Institute of High Pressure Physics (IHPP), Polish Academy of Sciences, Sokołowska 29/37, 01-142 Warsaw, Poland.

[2]Institute of Physics, Polish Academy of Sciences, aleja Lotników 32/46, 02-668 Warsaw, Poland.

[3]Łukasiewicz Research Network Institute of Microelectronics and Photonics, Aleja Lotników 32/46, 02-668 Warsaw, Poland.

*Corresponding author: sjs@unipress.waw.pl

### *Abstract*

We report the synthesis and characterization of a series of antimony-doped $SmFeAs_{1-x}Sb_xO_{0.8}F_{0.2}$ ($x$ = 0, 0.01, 0.03, 0.05, 0.1, 0.2, and 0.3) bulks to investigate the twin doping effects on the superconducting properties of SmFeAs(O,F) caused by fluorine (F) incorporation at O-site in SmO layer and antimony (Sb) substitution at As-site in the conducting layer (FeAs). Since the antimony (Sb) has a larger size than arsenic (As), the enhancement of lattice parameters has been confirmed by the XRD analysis. Microstructural analysis confirms that Sb-doping leads to a small improvement in the sample density and an increase in the inhomogeneity of the constituent elements, especially at higher Sb-doping levels. The parent compound $SmFeAsO_{0.8}F_{0.2}$ has shown the superconducting transition ($T_c$) at ~54 K, which is systematically reduced with the antimony doping contents ($x$). Our investigation indicates that the Sb-doped SmFeAs(O,F) phase at low levels is less prone to the multiphase formation than at high levels, which affects the inter- and intragranular behaviour originating from the microstructure nature of 1111 bulks. The critical current density ($J_c$) of the parent compound has almost the same value as previously reported, which is suppressed slowly with increased Sb-doping. It could be due to the reduced grain connections and the effective pinning centers. This study confirms that the superconducting FeAs layer doping with larger ions at arsenic sites does not support the superconducting properties of Sm1111, which is a distinct behavior from that of Sb-doped CeFeAs(O,F) and LaFeAs(O,F).

**Keywords:** Iron-based superconductor, 1111 family, polycrystalline samples, critical transition temperature, critical current density

# I. INTRODUCTION

Iron based superconductors (FBS) have many families, that are categorized based on the crystal structure of their parent compounds [1]. Among these, 1111 (*RE*FeAsO; *RE*1111; *RE* = rare earth) [2]-[4] family is a strong contender for high magnetic field applications and provides the highest transition temperature of 58 K for FBS. This family has a tetragonal ZrCuSiAs-type structure with an alternating layer of *RE*O and FeAs stacked along the *c*-axis [5],[6]. *RE*O and FeAs are called as charge reservoirs and conducting (superconducting) layers, respectively [1],[2]. The parent compound *RE*FeAsO does not show the superconductivity but exhibits the magnetic and structural transition at around 155 K [7]. Chemical doping generally plays an important role in inducing the superconductivity in this 1111 family due to the chemical pressure effect and by suppressing the magnetic and structural transitions [4]. Various kind of doping including F-doping at O-sites [2,4,11,12], Co/Mn/Ni at Fe sites [13,14,15], etc, are reported in which F doping play an effective role in enhancing the superconducting transition. Sm1111 shows the highest transition temperature of 58 K for F-doped Sm1111 ($SmFeAsO_{1-x}F_x$). We have reported previously the low-temperature synthesis of F-doped Sm1111 samples, which showed the highest transition temperature ($T_c$) of 58 K for 25% fluorine contents [5]. Generally, 20% F-doping is considered as an optimum doping level and shows a $T_c$ value of around 54 K [16]-[18]. There are many kinds of doping that have been reported for Sm1111, such as Co, Ni, Mn at Fe sites [19],[20], P doping at As sites [21]. However, there are no reports based on Sb-doping at arsenic (As) sites. Previous reports based on La1111 and Ce1111 show the enhancement of transition temperature by 2 K for the 5% antimony-doped $LaFeAsO_{0.8}F_{0.2}$ [22] and 5 K for the 10% Sb-doped $CeFeAsO_{0.9}F_{0.1}$ [7], respectively. These reports motivated us to study the larger ion radius effects, *i.e.*, antimony doping at arsenic sites in the superconducting FeAs layer of SmFeAs(O,F).

We have prepared a series of $SmFeAsO_{0.8}F_{0.2}$ bulks with various Sb-doping contents, *i.e.*, $SmFeAs_{1-x}Sb_xO_{0.8}F_{0.2}$ ($x$ = 0, 0.01, 0.03, 0.05, 0.1, 0.2 and 0.3) by one step solid-state reaction methods. These bulks are characterized by XRD, microstructure, transport, and magnetic measurements. The superconducting transition temperature ($T_c$) is reduced systematically with increasing Sb-doping levels. The calculated critical current density ($J_c$) for the parent compound is of the order of $10^3$ A/cm$^2$ at 0 T and 5 K, which is also decreased with the small amount of Sb-doping at arsenic sites, suggesting the reduced pinning centers. These investigations imply that the doping of the larger ionic size of Sb in the FeAs layer does not support the superconducting properties of F-doped Sm1111.

## II. EXPERIMETAL

Polycrystalline samples were prepared by using the high-purity precursors Sm powders (99.9%), As chunks (99.99%), Sb powder (99.9%), Fe powder (99.99%), $Fe_2O_3$ powder (99.85%) and $FeF_2$ powder (99%). First, SmAs was synthesized by reacting Sm and As powder at 500-550 °C for 15 hours, and SmSb was obtained by reacting Sm and Sb at 500°C for 24 hours. These raw materials were mixed according to the stoichiometric formula $SmFeAs_{1-x}Sb_xO_{0.8}F_{0.2}$ ($x$ = 0, 0.01, 0.03, 0.05, 0.1, 0.2, and 0.3), grounded, and compacted into pellets, which were placed in a tantalum tube as a crucible. This tube was inserted, sealed in an evacuated quartz tube, and annealed at 900°C for 45 hours. After this reaction, these tubes were opened inside the glove box. All chemical reaction processes were performed inside an inert gas glove box with very low ppm oxygen and moisture levels. After completing the reaction, the prepared bulks were stable in the open-air atmosphere. Structural analysis has been performed by X-ray diffraction using Cu-K$\alpha$ radiation. ICDD PDF4+2021 standard diffraction patterns database and Rigaku's PDXL software were applied to perform the profile analysis and the calculation of lattice parameters. Microstructural analysis and elemental mapping were done by Zeiss Ultra Plus field-emission scanning electron microscope equipped with energy-dispersive X-ray analysis (EDAX). Vibrating-sample magnetometry (VSM) attached to Quantum design PPMS was used for the magnetic measurements up to 9 T field. The temperature dependence of the resistivity measurements was carried out by a four-probe method attached to a closed-cycle refrigerator (CCR) in a zero magnetic field with a temperature range of 7-300 K.

## III. RESULTS AND DISCUSSION

Fig. 1(a) shows the X-ray diffraction (XRD) patterns for the nominal composition of $SmFeAs_{1-x}Sb_xO_{0.8}F_{0.2}$ bulks. The parent compound, $SmFeAsO_{0.8}F_{0.2}$ bulks, depicts the impurity phase of the SmOF phase, which is similar to the previous reports [5]. Interestingly, this impurity phase is almost constant for the various Sb-doping content; hence, it suggests that all samples have almost the same amount of fluorine. However, the $Fe_2Sb$ phase is observed at high Sb-doping contents. All samples have a tetragonal phase with the space group of *P4/nmm*. The obtained lattice parameters are $a$ = 3.928(7) Å and $c$ = 8.497(9) Å for $x$ = 0, $a$ = 3.930(9) Å and $c$ = 8.503(3) Å for $x$ = 0.03, and $a$ = 3.940(1) Å and $c$ = 8.540(9) Å for $x$ = 0.3. The variation of the lattice parameter '$c$' with the Sb-doping level (*x*) is shown in Fig. 1(b) for our Sb-doped $SmFeAsO_{0.8}F_{0.2}$ and also the reported Sb-doped $CeFeAsO_{0.8}F_{0.2}$. A systematic increase of the lattice parameters '*c*' is observed for Sb-doped $SmFeAsO_{0.8}F_{0.2}$ compared to that of the parent compound ($x = 0$) due to the large size of the Sb-ion relative to the As-ion. This behavior of the lattice parameter is well agreed with the reported Sb-doped CeFeAs(O,F) [7] as shown in Fig. 1(b), and also with Sb-doped LaFeAs(O,F) [22]. Hence, it confirms that Sb contents were successfully inserted into the lattice of F-doped SmFeAsO.

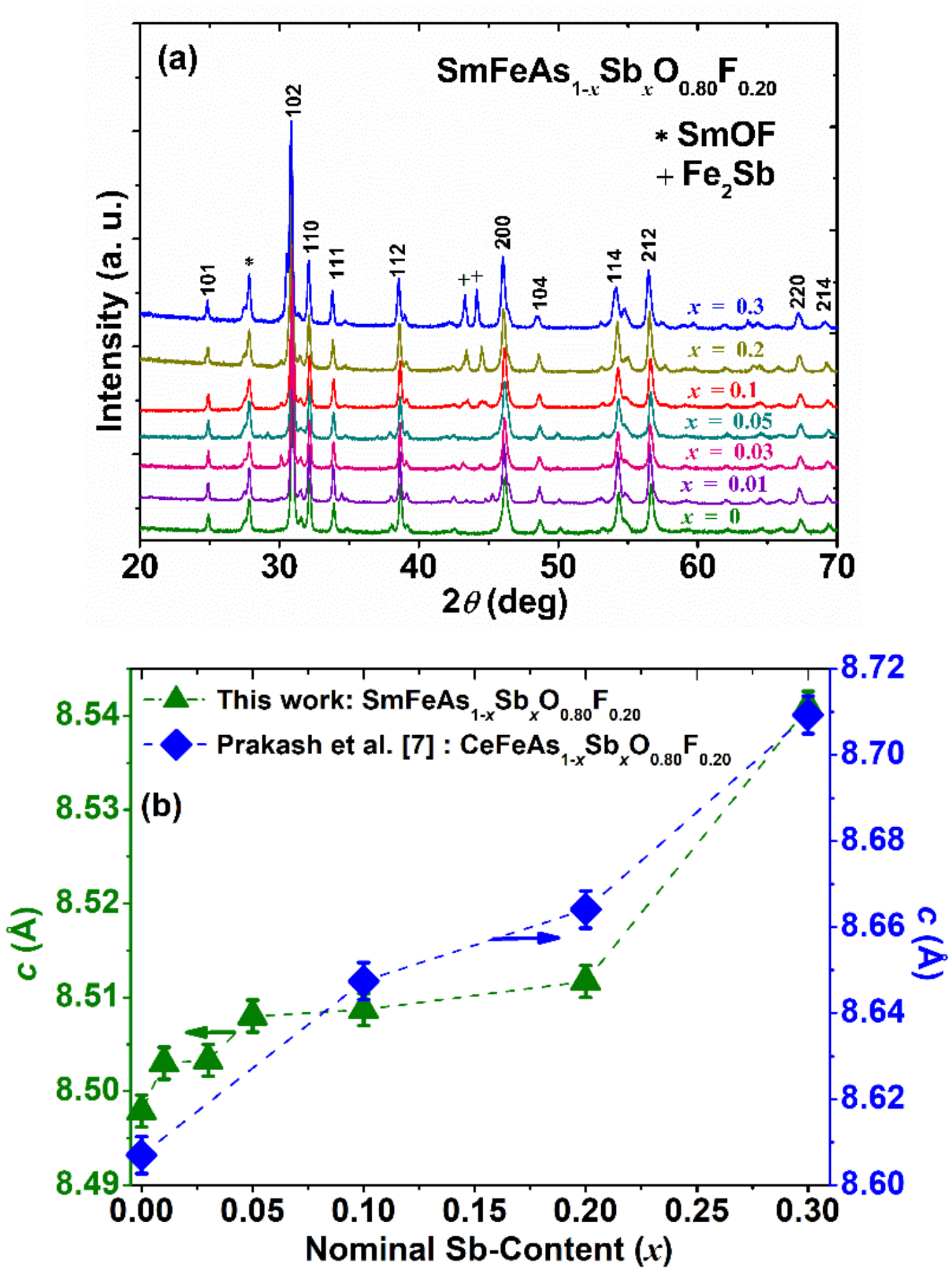

**Fig. 1. (a)** Powder X-ray diffraction patterns of various Sb-doped Sm1111 ($SmFeAs_{1-x}Sb_xO_{0.8}F_{0.2}$) bulks **(b)** The variation of lattice parameter '*c*' with respect to the nominal Sb-content (*x*). The arrow used guides the eye to the vertical scale for the respective graph.

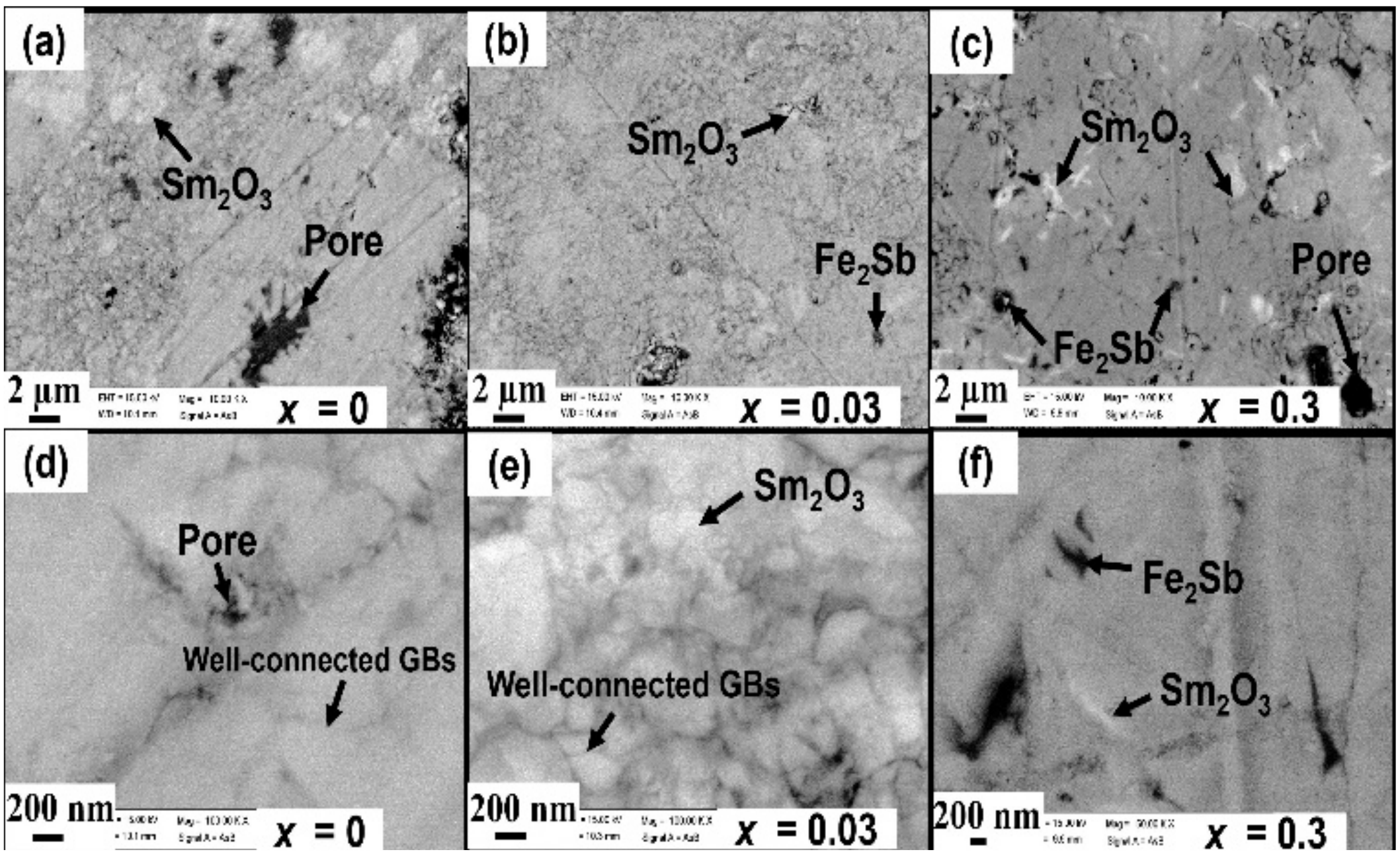

**Fig. 2.** Back-scattered images (BSE) of $SmFeAs_{1-x}Sb_xO_{0.8}F_{0.2}$ bulks: **(a),(d)** for $x = 0$; **(b),(e)** for $x = 0.03$ and **(c), (f)** for $x = 0.3$.

In order to understand the microstructural analysis, the backscattered images (BSE) of these samples are shown in Fig. 2 for the parent and Sb-doped Sm1111 bulks, whereas the elemental mappings are added as Fig. S1 in the supplementary file. White, gray, and black contrasts are observed for $Sm_2O_3$ (SmOF), $SmFeAsO_{0.8}F_{0.2}$, and $Fe_2Sb$ or pores, respectively. $Sm_2O_3$ and pores are observed for the parent compound with the main Sm1111 phase and many well-connected grains do exist. Low Sb-doping levels such as $x = 0.03$, have reduced the number of pores, and $Fe_2Sb$ is also observed in a few places with $Sm_2O_3$ (SmOF) (Fig. 2(b) and 2(e)). With further increase of Sb-doping, the impurity phase of $Fe_2Sb$ is increased and observed at many places, as shown in Figs. 2(c) and 2(f) for $x = 0.3$. Sb-doped samples appear denser than the parent compound, which could be due to the presence of the impurity phase of $Fe_2Sb$. This impurity is observed both inside and between superconducting grains, hence, it suggests that the intergrain connections are reduced with Sb doping ($x$). Fig. S1(i) depicts an almost homogeneous distribution of various elements for the parent compounds, but in a few places, SmOF is observed, as similar to XRD measurements. For very low Sb-doped samples, such as $x = 0.01$, the elemental mappings are very similar to the parent. The inhomogeneity of the constituent elements is more visible with increasing Sb-content, as depicted in Figs. S1(iii), (iv), and (v). By considering the theoretical density of 7.1 g/cm$^3$ for Sm1111 [5], the density for $x = 0$, 0.01, 0.03, 0.05, and 0.2 is around 37%, 42%, 45%, 48%, and 44%, which suggests a slight improvement in the density of the superconducting phase, but concurrently, the impurity phases were also increased especially at higher Sb-doping contents, as observed from the microstructural and XRD analysis.

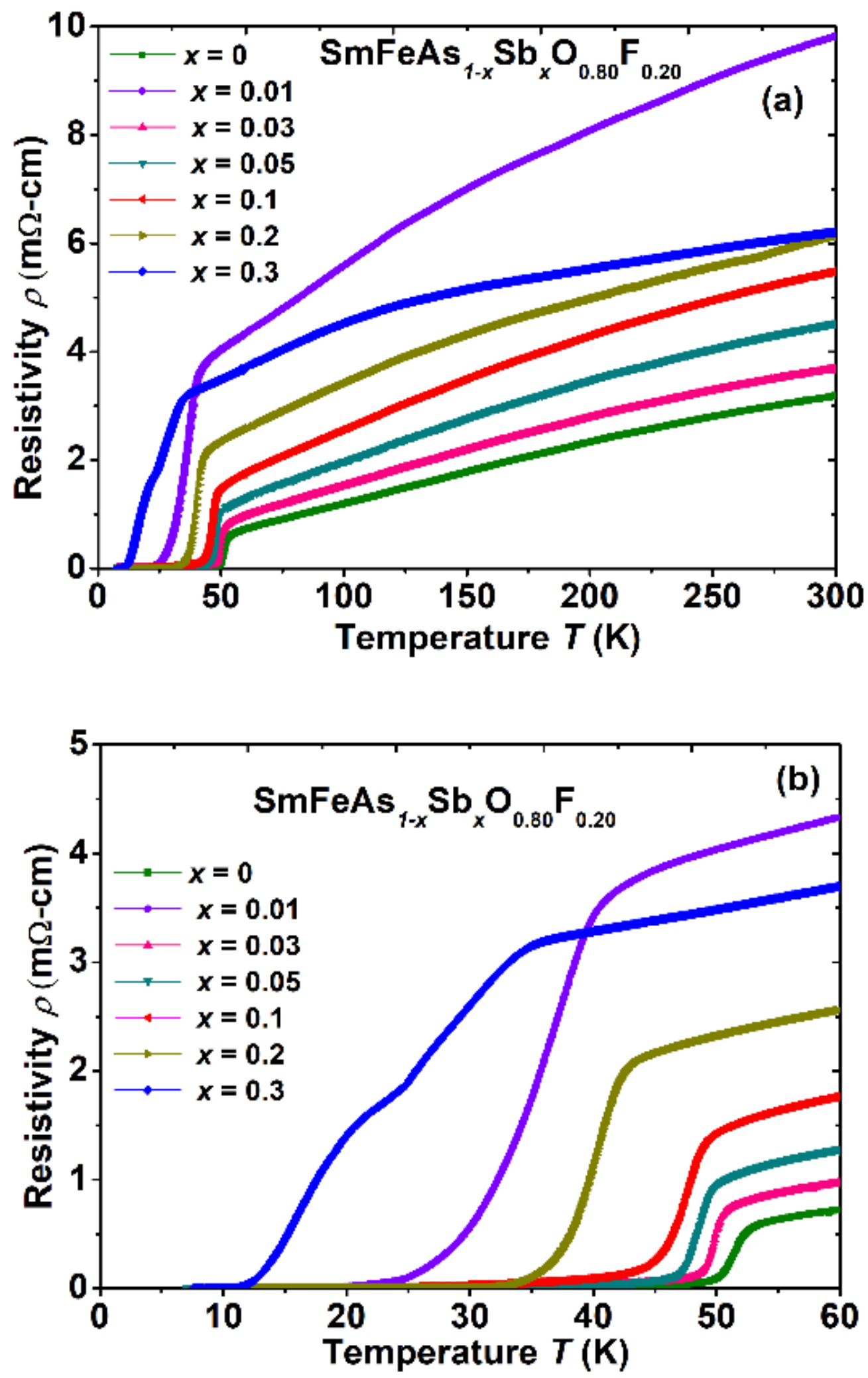

**Fig. 3. (a)** The variation of resistivity with temperature in zero magnetic field **(b)** Low-temperature variation of the resistivity with temperature parameter for $SmFeAs_{1-x}Sb_xO_{0.8}F_{0.2}$ bulks.

The temperature variation of the resistivity ($\rho$) at zero magnetic field is illustrated in Fig. 3(a) up to room temperature, which decreases monotonically with temperature and confirms the superconductivity for all samples below 55 K. The normal state resistivity of the parent compound is increased with the Sb-doping contents, which could be due to the impurity phase of $Fe_2Sb$, as confirmed by the XRD and microstructural analysis. A very small amount of Sb addition, *i.e.*, $x$ = 0.01, has shown a higher normal state resistivity value compared to other samples, which could be possible because of the inhomogeneous distribution of Sb. To confirm the behavior of $x$ = 0.01, this sample was reproduced in 2-3 batches under the same growth conditions, and its resistivity behavior is displayed in the supplementary file as Fig. S2, which depicts a significant reduction of $T_c$ as observed for $x$ = 0.01 sample in Fig. 3(a). The low-temperature resistivity behavior for all these samples is shown in Fig. 3(b). The parent compound depicts the onset transition temperature up to 53.8 K, whereas this value is decreased with Sb-doping contents systematically except the sample $x$ = 0.01, which agrees with the variation of lattice parameters observed from XRD data and suggests that Sb content is incorporated into the superconducting lattice (Fig. 1(b)). A very low amount of Sb doping *i.e.* $x$ = 0.01, has a high resistivity behavior and a huge reduction of the onset $T_c$ (~41 K) compared to other Sb-doped samples.

To confirm the Meissner effect of these bulks, the diamagnetic signal in zero-field-cooled (ZFC) and field-cooled (FC) magnetization modes was measured in the temperature range from 5 to 60 K at a magnetic field of 20 Oe. The normalized magnetization by the lowest magnetic moment measured at 5 K is depicted in Fig. 4(a) for $x$ = 0, 0.01, 0.03, 0.05, and $x$ = 0.2. The negative magnetic response in field-cooled data represents the trapped magnetic flux inside the bulk samples. These diamagnetic signals confirm the superconductivity of these samples. Interestingly, the $T_c$ is observed around 42-43 K for $x$ =0.01, which is slightly higher than that of resistivity measurement. It suggests that there was not a homogenous distribution of Sb contents due to a very small amount of Sb contents (Fig. S3). Further Sb-doped samples have depicted a systematic decrease of transition temperature with Sb contents as similar to the resistivity behaviors. The two-step behaviors of the parent compound are similar to the previous reports [5] due to a weak-link behavior as well known for FBS [2,23]. It is not observed for Sb doped samples which could be possible due to the improvement of the sample density as discussed with the microstructural analysis. The onset transition temperature ($T_c^{onset}$) for $SmFeAsO_{0.8}F_{0.2}$ is observed around 53 K, whereas the addition of Sb-doping has improved the diamagnetic transition from a two-step to a single-step transition, and a $T_c^{onset}$ of 49 K, 48.7 K, and 42 K is observed for $x$ = 0.03, 0.05 and 0.2. This single-step transition can be interpreted with the intragranular properties of the polycrystalline samples, as discussed for other FBS [2],[23]. The reduction of transition temperature resulted from the decrease of chemical pressure in the superconducting 1111 lattice due to the larger size of Sb dopants [7,22].

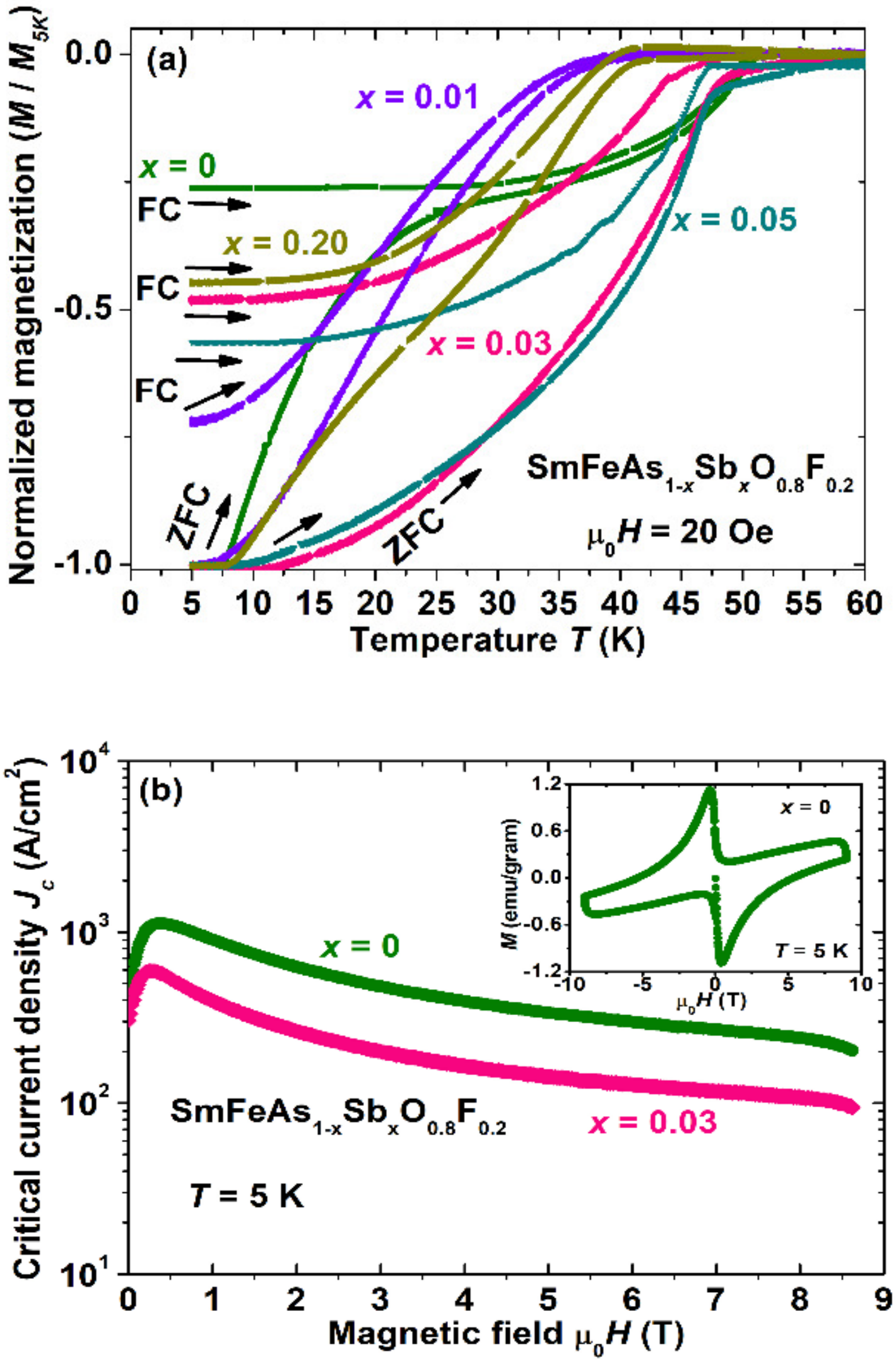

**Fig. 4. (a)** The temperature dependence of the normalized magnetic moment ($M$ / $M_{5\ K}$) at 20 Oe magnetic field. **(b)** The magnetic field dependence of the critical current density $J_c$ for the parent compound ($x = 0$) and $x = 0.03$. The inset of Fig. 4(b) depicts the magnetic hysteresis loop ($M$-$H$) at 5 K for the sample $x = 0$, where $M$ is the magnetic moment.

At the temperature of 5 K, the magnetic hysteresis loop ($M$-$H$) data is collected for the parent compound ($x = 0$) and 3% Sb- added sample ($x = 0.03$) measured in the magnetic field up to 9 T. The inset of Fig. 4(b) depicts the magnetic hysteresis loop for the parent compound, which has a small-width hysteresis loop due to weak intragranular pinning or weak-link behaviors, which are well-known behaviors reported for oxypnictide superconductors [2],[23]. Similar hysteresis loops, here, are also observed for other Sb-doped Sm1111. The critical current density ($J_c$) is calculated by considering the Bean critical state model by using these isothermal magnetic *M-H* loops, *i.e.*, $J_c = 20\Delta m/Va(1 - a/3b)$ [24], where $V$ is the volume of the polycrystalline sample, $a$ and $b$ are shorter and longer edges ($a < b$) of the sample, and $\Delta m$ is the hysteresis loop width obtained from the measured *M-H* loop. The rectangular-shaped samples were used for the hysteresis loop measurements. Assuming that supercurrents flow at a density equal to the critical-current density $J_c(H)$, which is constant regardless of the magnetic field ($H$), the Bean model estimates the $J_c$ of a superconductor. The magnetic behavior of the parent compound, as shown in Fig. 4(b), is consistent with the previous report [5] and its maximum $J_c$ value is of the order of $10^3$ A/cm$^2$ at 5 K and 0.5 T, as reported [5]. The sample with $x = 0.03$ has a decreased critical current density, which might be due to the reduced pinning centers or weak grain connections due to the presence of impurity phases, as discussed elsewhere for other FBS [1],[2],[6].

To summarize our main findings, the variation of onset transition temperature ($T_c$) and the residual resistivity ratio ($RRR = \rho_{300K} / \rho_{55K}$) with antimony doping contents ($x$) are depicted in Fig. 5(a) and 5(b). The transition temperature is decreased with antimony doping at As-sites from 53.8 K as obtained for the parent compounds ($x = 0$). This decreased $T_c$ is related to the reduction of chemical pressure due to the larger size of Sb dopants [1],[2]. The variation of *RRR* is shown in Fig. 5(b), which generally represents the homogeneity, better grain connections, and a phase-pure sample. The parent compound has an *RRR* of 5, which is similar to the reported value for F-doped Sm1111 bulks [5]. With Sb doping, a systematic reduction in *RRR* is observed, suggesting the enhancement of the inhomogeneity of these bulks, which is also confirmed by other measurements as discussed above. The inhomogeneity of the samples reduced the *RRR* value to 1.5 for 30% Sb-doped $SmFeAsO_{0.8}F_{0.2}$ samples ($x = 0.3$), which is well in agreement with the elemental mapping of these samples (Fig. S1). These results are completely different from those for Sb-doped LaFeAs(O,F) [22] and CeFeAs(O,F) [7] samples, where Sb doping enhanced the transition temperature by 2 to 5 K. However, Sb-doped $SmFeAsO_{0.8}F_{0.2}$ clearly demonstrates that the disorder in the conducting FeAs layers could not support the enhancement of the superconducting transition temperature in this multiband superconductor.

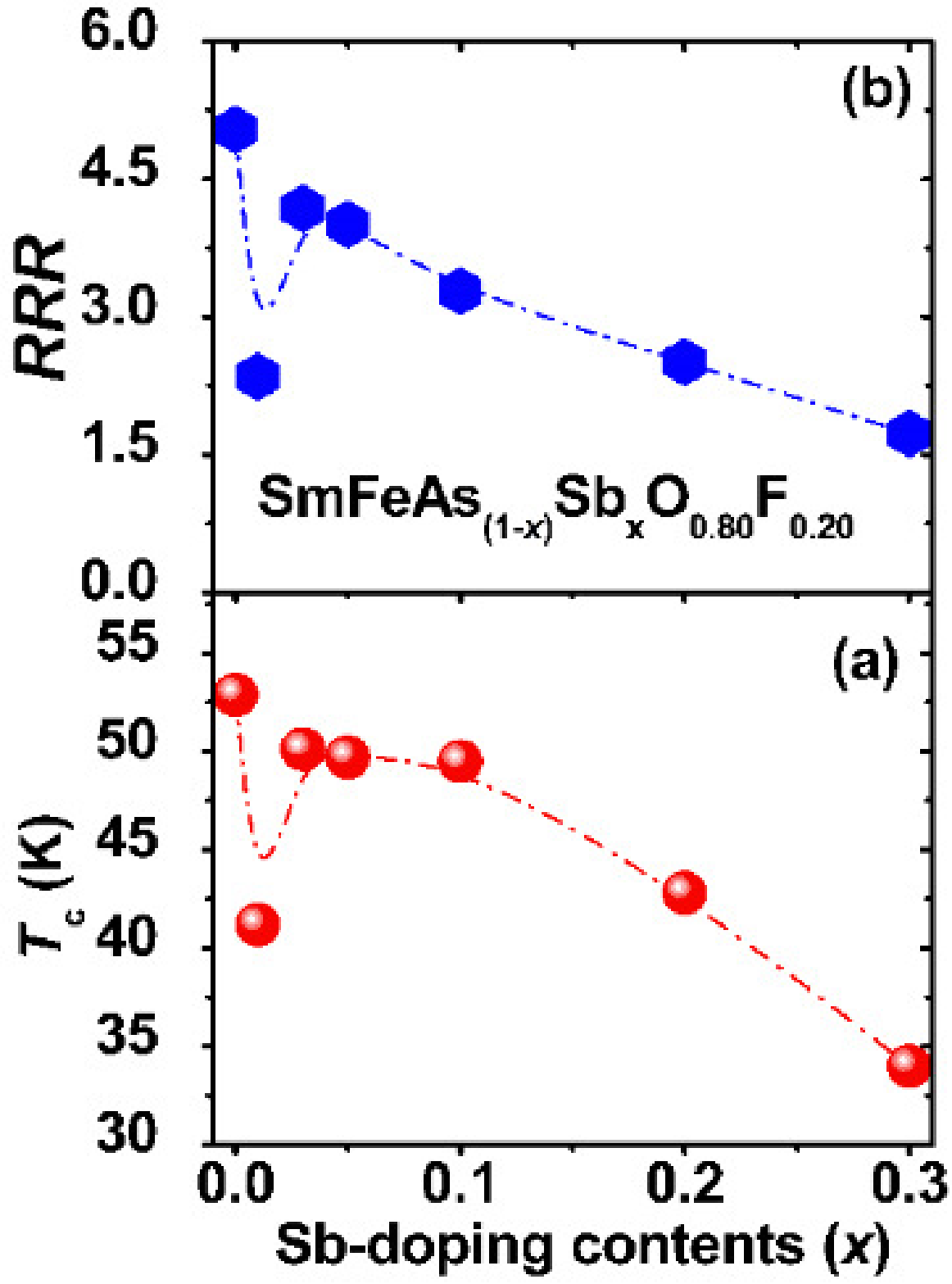

**Fig. 5. (a)** The onset transition temperature $T_c$ and **(b)** residual resistivity ratio ($RRR = \rho_{300K} / \rho_{55K}$) variation with Sb-doping contents ($x$) for $SmFeAs_{1-x}Sb_xO_{0.8}F_{0.2}$ bulks.

## IV. CONCLUSIONS

The antimony (Sb) doping effect at arsenic sites has been studied by preparing a series of samples using a low-temperature synthesis method. The superconducting properties of $SmFeAsO_{0.8}F_{0.2}$ showed a transition temperature of 54 K and a critical current density of the order of $10^3$ A/cm$^2$ at 0 T, 5 K, which is almost the same as previously reported. Microstructural and XRD analysis depicts the enhancement of impurity phases at a high amount of Sb-doping. Additionally, the large size of Sb causes an increase in the lattice parameters, which suggests a reduction of the chemical pressure inside the superconducting tetragonal lattice. As a result, the antimony substitution reduces the superconducting transition in a systematic manner with the exception of a very low doping level ($x$ = 0.01). In addition, the critical current density is also decreased for Sb-doped samples compared to the parent compound ($x = 0$). This may be attributed to a reduction in the intergrain connections and effective pinning centers. Our studies confirm that the disorder caused by the larger size of dopants is not suitable for the improvement of the superconducting properties of SmFeAs(O,F) compared to the previous reports based on Sb-doped CeFeAs(O,F) and LaFeAs(O,F). Our results suggest further studies on antimony-doped oxypnictide superconductors appear to be an interesting area of research.

## V. ACKNOWLEDGMENTS

This research was funded by National Science Centre (NCN), Poland, grant number “2021/42/E/ST5/00262” (SONATA-BIS 11). S.J.S. acknowledges financial support from National Science Centre (NCN), Poland through research Project number: 2021/42/E/ST5/00262.

## REFERENCES

[1] H. Hosono, A. Yamamoto, H. Hiramatsu, and Y. Ma, "Recent advances in iron-based superconductors toward applications," *Materials Today*, vol. 21, no. 3, pp. 278–302, Apr. 01, 2018. doi: 10.1016/j.mattod.2017.09.006.

[2] S. J. Singh and M. I. Sturza, "Bulk and single crystal growth progress of iron-based superconductors (Fbs): 1111 and 1144," *Crystals*, vol. 12, no. 1, pp. 20, Jan. 01, 2022. doi: 10.3390/cryst12010020.

[3] N. Fujiwara and H. Takahashi, "Electric and magnetic properties of $LaFeAsO_{1-x}H_x$ under pressure," *Jpn. J. Appl*, vol. 56, no. 5, pp. 05FA08, May 01, 2017. doi: 10.7567/JJAP.56.05FA08.

[4] Y. Kamihara, T. Watanabe, M. Hirano, and H. Hosono, "Iron-based layered superconductor $LaO_{1-x}F_xFeAs$ ($x$ = 0.05-0.12) with Tc = 26 K," *J. Am. Chem. Soc.*, vol. 130, no. 11, pp. 3296–3297, Mar. 2008, doi: 10.1021/ja800073m.

[5] S. J. Singh, J. Shimoyama, A. Yamamoto, H. Ogino, and K. Kishio, "Transition temperature and upper critical field in $SmFeAsO_{1-x}F_x$ synthesized at low heating temperatures," *IEEE Trans. on App. Super.*, vol. 23, no. 3, pp. 7300605, June 2013, doi: 10.1109/TASC.2013.2239352.

[6] J. I. Shimoyama, "Potentials of iron-based superconductors for practical future materials," *Supercond. Sci. Technol.*, vol. 27, no. 4, pp. 044002, 2014, doi: 10.1088/0953-2048/27/4/044002.

[7] J. Prakash, S. J. Singh, G. Thakur, S. Patnaik, and A. K. Ganguli, "The effect of antimony doping on the transport and magnetic properties of Ce(O/F)FeAs," *Supercond. Sci Technol.*, vol. 24, no. 12, pp. 125008, Dec. 2011, doi: 10.1088/0953-2048/24/12/125008.

[8] Q. Si, R. Yu, and E. Abrahams, "High-temperature superconductivity in iron pnictides and chalcogenides," *Nature Reviews Materials*, Nature Publishing Group, vol. 1, pp. 16017, Mar. 11, 2016. doi: 10.1038/natrevmats.2016.17.

[9] J. A. Rodgers *et al.*, "Suppression of the superconducting transition of $RFeAsO_{1-x}F_x$ (R=Tb, Dy, and Ho)," *Phys Rev B Condens Matter Mater Phys*, vol. 80, no. 5, pp.052508, Aug. 2009, doi: 10.1103/PhysRevB.80.052508.

[10] C. J. Court and J. M. Cole, “Magnetic and superconducting phase diagrams and transition temperatures predicted using text mining and machine learning,” *NPJ Comput Mater*, vol. 6, no. 1, pp. 18, Dec. 2020, doi: 10.1038/s41524-020-0287-8.

[11] P. Cheng *et al.*, “Superconductivity at 36 K in gadolinium-arsenide oxides $GdO_{1-x}F_xFeAs$,” *Sci. China: Phys. Mech. Astron.*, vol. 51, no. 6, pp. 719–722, Jun. 2008, doi: 10.1007/s11433-008-0075-9.

[12] R. Prozorov, M. E. Tillman, E. D. Mun, and P. C. Canfield, “Intrinsic magnetic properties of the superconductor $NdFeAsO_{0.9}F_{0.1}$ from local and global measurements,” *New J Phys*, vol. 11, pp. 035004, Mar. 2009, doi: 10.1088/1367-2630/11/3/035004.

[13] A. Marcinkova, D. A. M. Grist, I. Margiolaki, T. C. Hansen, S. Margadonna, and J. W. G. Bos, “Superconductivity in $NdFe_{1-x}CoxAsO$ (0.05<x<0.20 ) and rare-earth magnetic ordering in NdCoAsO,” *Phys. Rev. B*, vol. 81, no. 6, pp. 064511, Feb. 2010, doi: 10.1103/PhysRevB.81.064511.

[14] S. Matsuishi, Y. Inoue, T. Nomura, H. Yanagi, M. Hirano, and H. Hosono, “Superconductivity induced by co-doping in quaternary fluoroarsenide CaFeAsF,” *J Am Chem Soc*, vol. 130, no. 44, pp. 14428–14429, Nov. 2008, doi: 10.1021/ja806357j.

[15] S. J. Singh, J. Prakash, A. Pal, S. Patnaik, V. P. S. Awana, and A. K. Ganguli, “Study of Ni and Zn doped CeOFeAs: Effect on the structural transition and specific heat capacity,” *Physica C: Superconductivity and its Applications*, vol. 490, pp. 49–54, 2013, doi: 10.1016/j.physc.2013.04.077.

[16] J. B. Anooja, P. M. Aswathy, P. M. Sarun, and U. Syamaprasad, “A novel low temperature synthesis route for $SmFeAsO_{1-x}F_x$ bulk superconductor with improved transport properties,” *Journal of Alloys and Compounds*, vol. 514. Elsevier Ltd, pp. 1–5, Feb. 15, 2012. doi: 10.1016/j.jallcom.2011.11.041.

[17] C. Wang *et al.*, “Low-temperature synthesis of SmO0.8F0.2FeAs superconductor with Tc = 56.1K,” *Supercond. Sci. Technol.*, vol. 23, no. 5, pp. 055002, Apr. 01, 2010, doi: 10.1088/0953-2048/23/5/055002.

[18] Y. Ding *et al.*, “Density effect on critical current density and flux pinning properties of polycrystalline $SmFeAsO_{1-x}F_x$ superconductor,” *Supercond Sci Technol*, vol. 24, no. 12, pp. 125012, Dec. 2011, doi: 10.1088/0953-2048/24/12/125012.

[19] Y. Qi, Z. Gao, L. Wang, D. Wang, X. Zhang, and Y. Ma, "Superconductivity in co-doped SmFeAsO," *Supercond. Sci. Technol.*, vol. 21, no. 11, pp. 115016, Nov. 2008, doi: 10.1088/0953-2048/21/11/115016.

[20] S. J. Singh, J. Shimoyama, A. Yamamoto, H. Ogino, and K. Kishio, "Effects of Mn and Ni doping on the superconductivity of SmFeAs(O,F)," Phys. C: Supercond. Appl., vol. 494, pp. 57–61, 2013, doi: 10.1016/j.physc.2013.04.045.

[21] S. J. Singh, J.-I. Shimoyama, A. Yamamoto, H. Ogino, and K. Kishio, "Effects of phosphorous doping on the superconducting properties of SmFeAs(O,F)," *Physica C*, vol. 504, pp. 19-23, Sep. 2014, https://doi.org/10.1016/j.physc.2014.04.022

[22] S. J. Singh, J. Prakash, S. Patnaik, and A. K. Ganguli, "Enhancement of the superconducting transition temperature and upper critical field of $LaO_{0.8}F0._{2}FeAs$ with antimony doping," *Supercond Sci Technol*, vol. 22, no. 4, pp. 045017, 2009, doi: 10.1088/0953-2048/22/4/045017.

[23] A. Yamamoto *et al.*, "Evidence for electromagnetic granularity in polycrystalline Sm1111 iron-pnictides with enhanced phase purity," *Supercond. Sci. Technol.*, vol. 24, no. 4, pp. 045010, Apr. 2011, doi: 10.1088/0953-2048/24/4/045010.

[24] C. P. Bean, "Magnetization of high-field superconductors." *Rev. Mod. Phys.*, vol. 36, pp. 31-39, 1964, doi: 10.1103/RevModPhys.36.31

# Supplementary Data

## Antimony Doping Effect on the Superconducting Properties of SmFeAs(O,F)

Mohammad Azam[1], Manasa Manasa[1], Tatiana Zajarniuk[2], Ryszard Diduszko[3], Tomasz Cetner[1], Andrzej Morawski[1], Andrzej Wiśniewski[2], Shiv J. Singh[1*]

[1]Institute of High Pressure Physics (IHPP), Polish Academy of Sciences, Sokołowska 29/37, 01-142 Warsaw, Poland.
[2]Institute of Physics, Polish Academy of Sciences, aleja Lotników 32/46, 02-668 Warsaw, Poland.

[3]Łukasiewicz Research Network Institute of Microelectronics and Photonics, Aleja Lotników 32/46, 02-668 Warsaw, Poland.

*Corresponding author: sjs@unipress.waw.pl

(i)

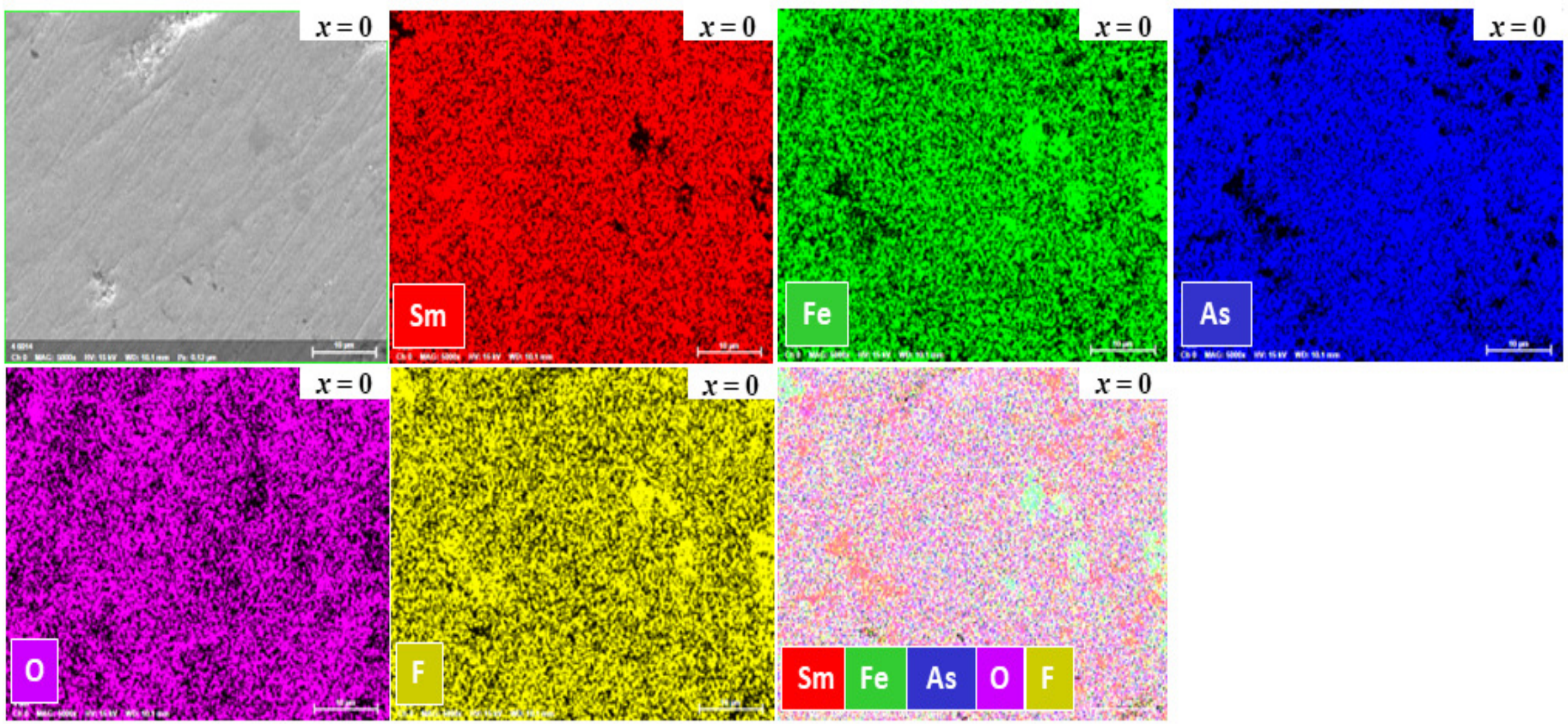

(ii)

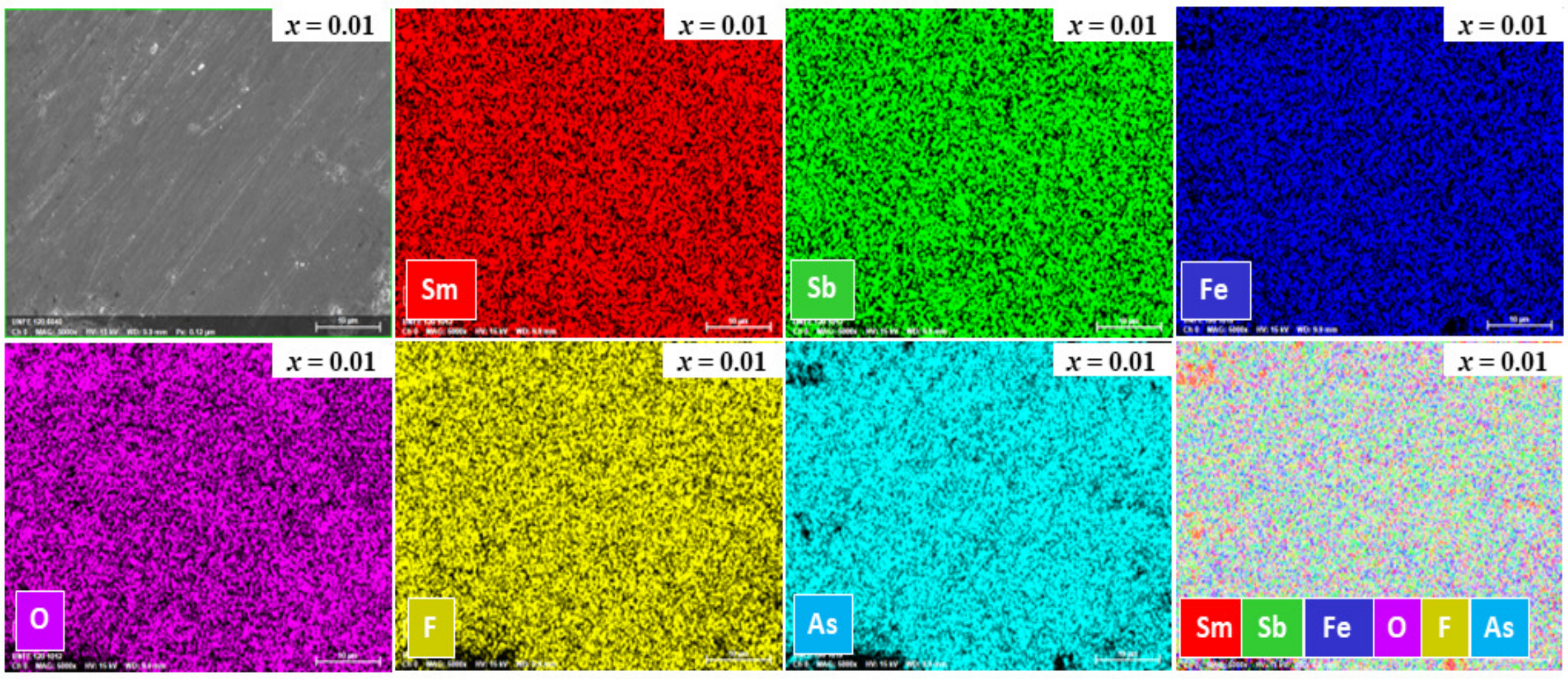

(iii)

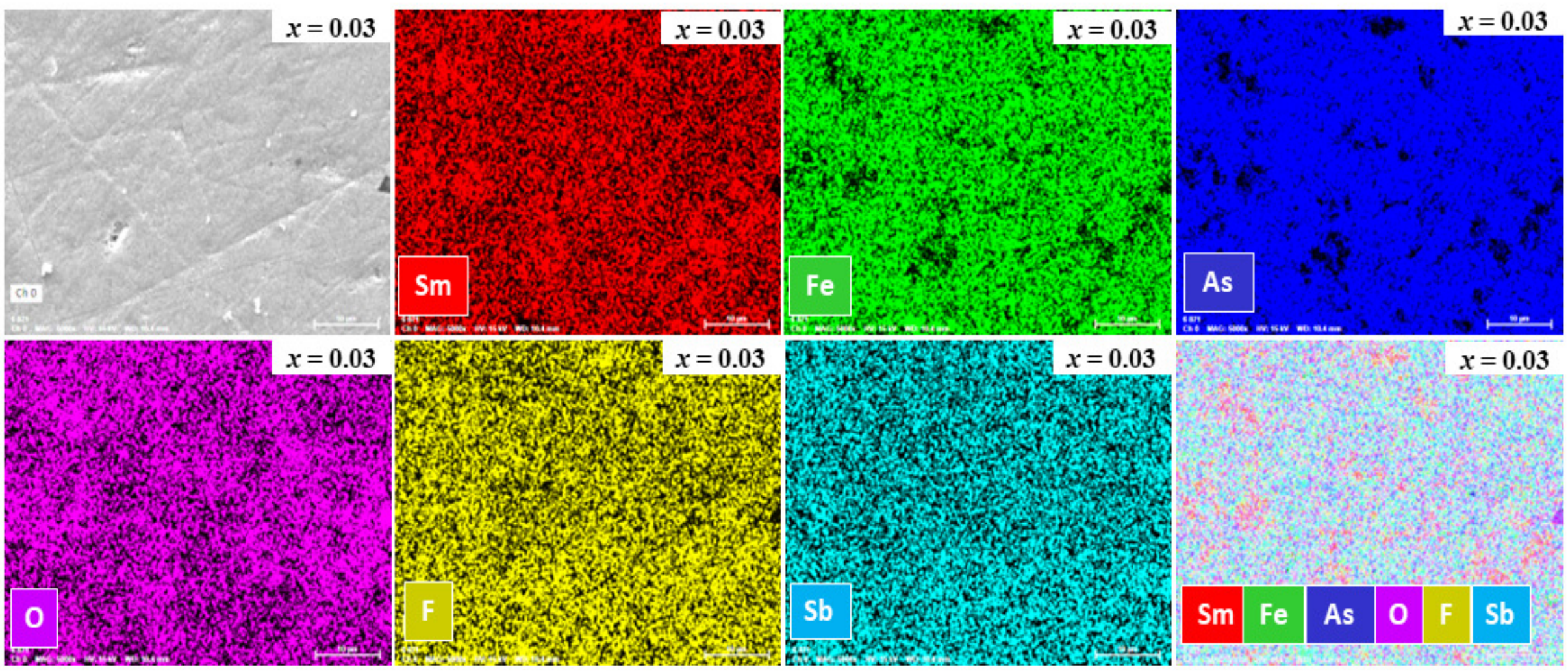

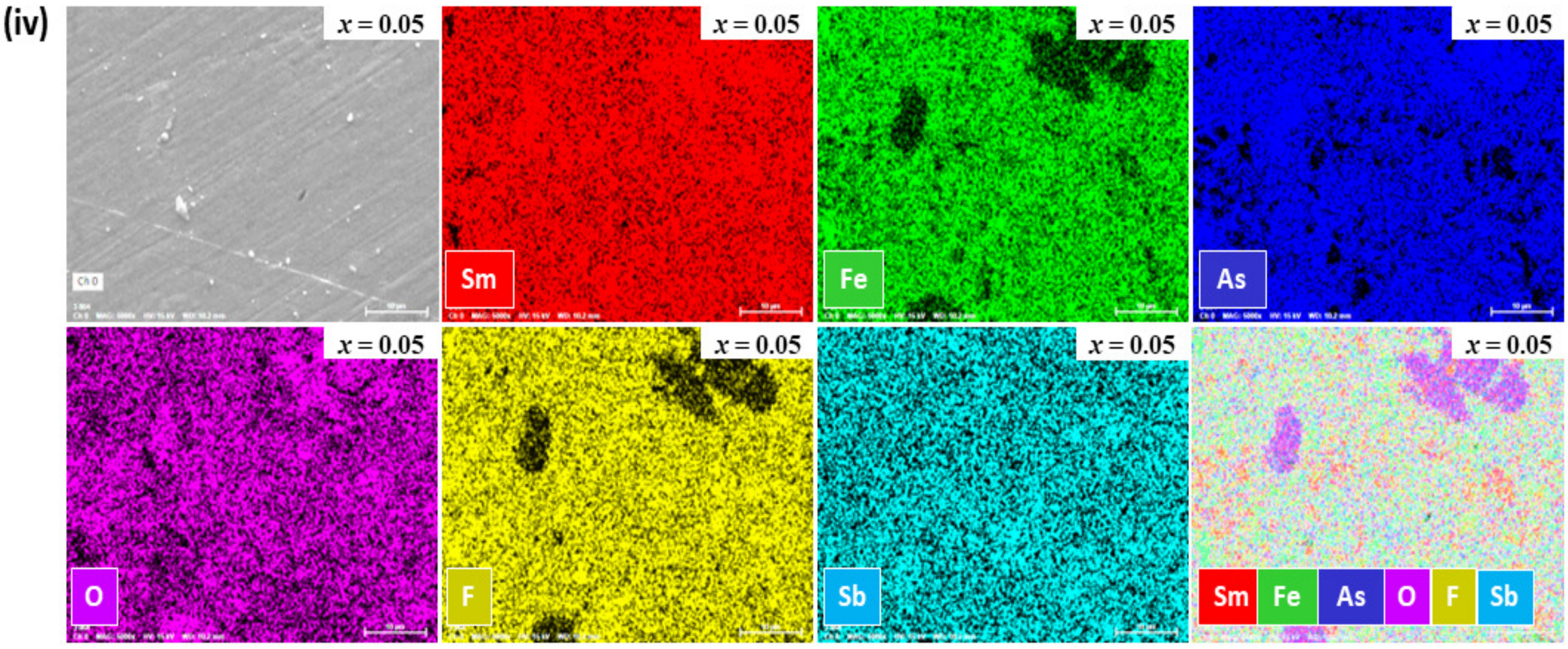

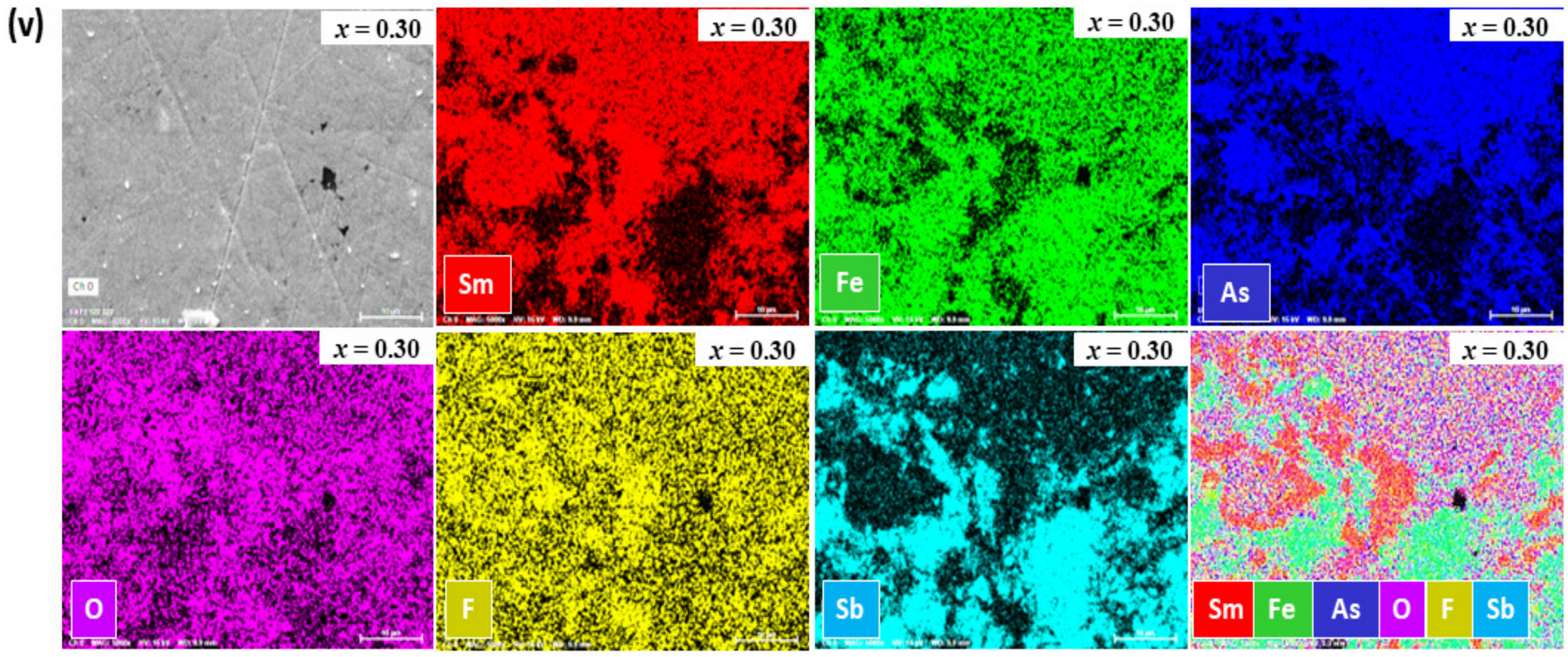

**Fig. S1.** Mapping of all constituent elements for $SmFeAs_{1-x}Sb_xO_{0.8}F_{0.2}$ **(i) f**or $x = 0$, **(ii)** for $x = 0.01$, **(iii)** for $x = 0.03$, **(iv)** for $x = 0.05$ and **(v)** for $x = 0.3$.

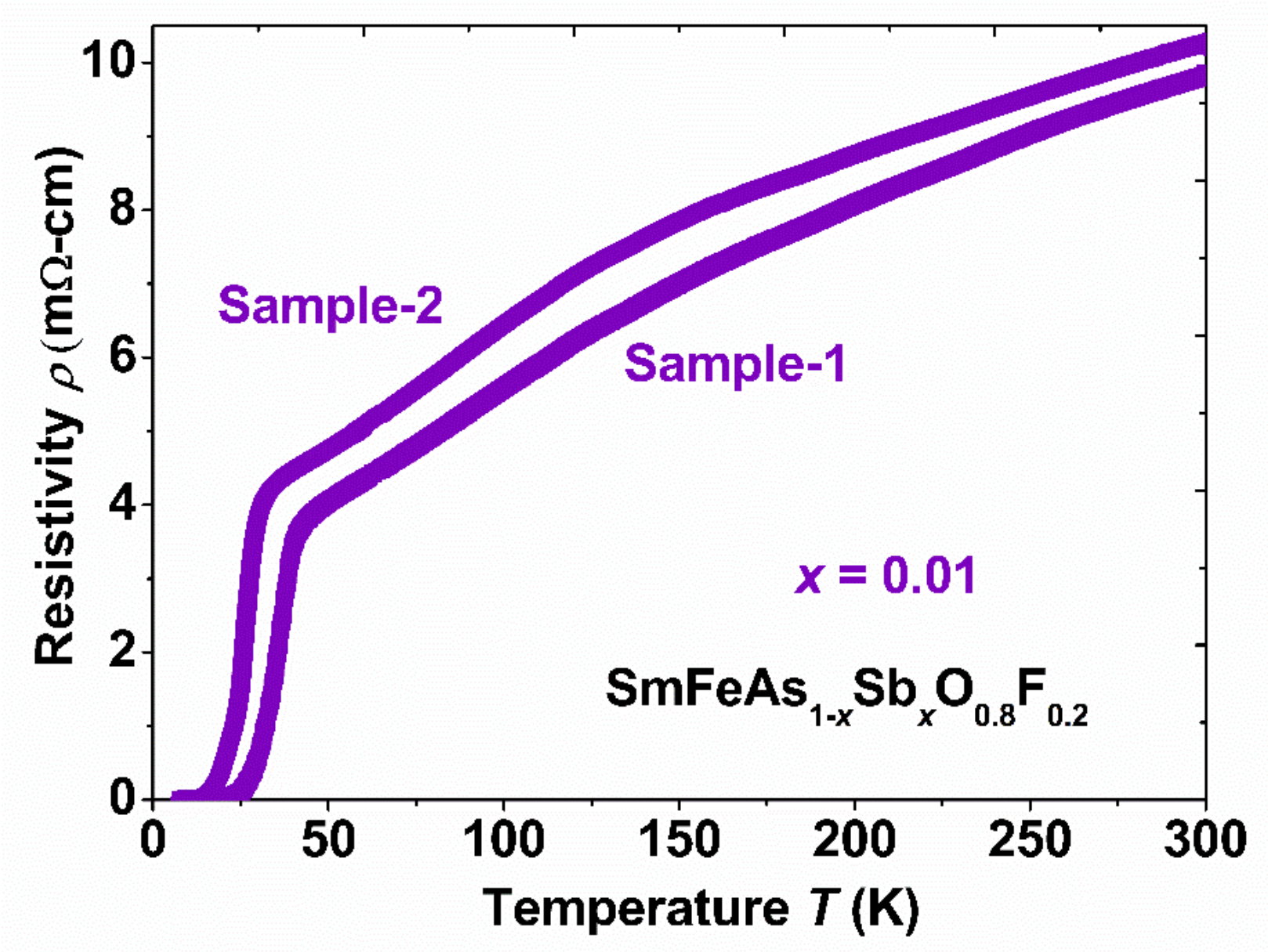

**Fig. S2.** The temperature dependence of resistivity behavior of $SmFeAs_{0.99}Sb_{.01}O_{0.8}F_{0.2}$ for sample-1 (batch-I) and sample-2 (batch-II).

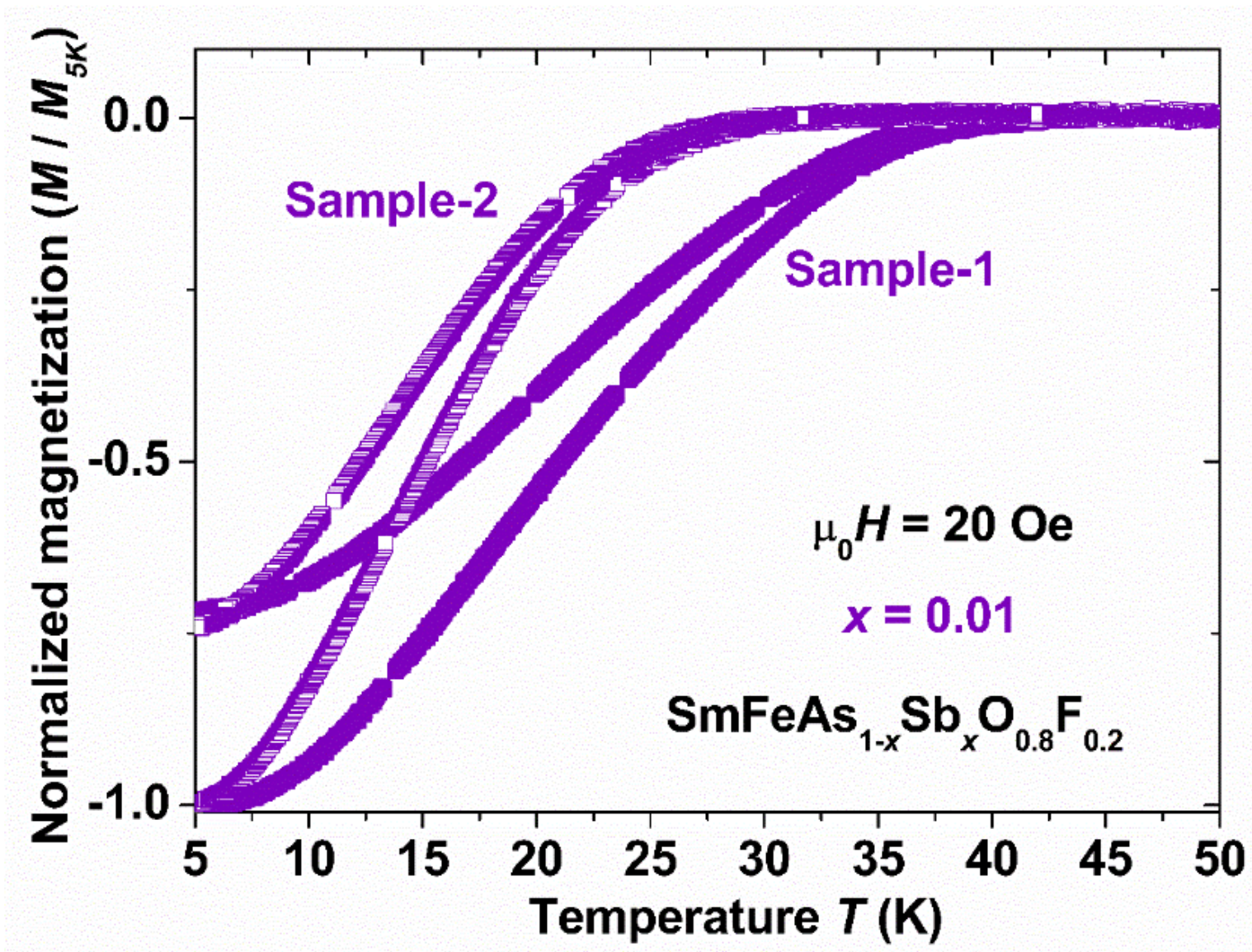

**Fig. S3.** The temperature dependence of the normalized magnetic moment ($M$ / $M_{5K}$) at 20 Oe magnetic field for $SmFeAs_{0.99}Sb_{0.01}O_{0.8}F_{0.2}$ for sample-1 (batch-I) and sample-2 (batch-II).